\documentclass{emulateapj}

\newcommand{\spitzer}{{\emph{Spitzer}}}

\newcommand{\kms}{\mbox{\,km\,s$^{-1}$}}

\newcommand{\lumin}{\mbox{\,erg~s$^{-1}$}}
\newcommand{\persec}{\mbox{\,s$^{-1}$}}

\newcommand{\CIV}{\ion{C}{4}}

\newcommand{\delg}{$\Delta(g-i)$}

\newcommand{\altcite}{\citealt}

\newcommand{\lir}{$l_{\rm ir}$}
\newcommand{\lopt}{$l_{\rm opt}$}
\newcommand{\Lir}{$L_{\rm ir}$}
\newcommand{\Lopt}{$L_{\rm opt}$}
\newcommand{\aopt}{$\alpha_{\nu{\rm,opt}}$}
\newcommand{\air}{$\alpha_{\nu{\rm,ir}}$}
\newcommand{\lthree}{$l_{3\mu{\rm m}}$}
\newcommand{\lfiveoh}{$l_{5000{\rm \AA}}$}
\newcommand{\Lthree}{$L_{3\mu{\rm m}}$}
\newcommand{\Lfiveoh}{$L_{5000{\rm \AA}}$}

\newcommand{\leightair}{$l_{8,\alpha}$}
\newcommand{\leightsed}{$l_{\rm 8,SED}$}
\newcommand{\Leight}{$L_{8\mu{\rm m}}$}
\newcommand{\Leightair}{$L_{8,\alpha}$}
\newcommand{\Leightsed}{$L_{\rm 8,SED}$}
\newcommand{\total}{234}
\newcommand{\subsample}{55}
\newcommand{\microns}{\micron}
\begin{document}
 
 
\shortauthors{Gallagher et al.}  
\shorttitle{Luminosity Effects on Mid-IR Quasar SEDs} 

\title{An Investigation into the Effects of Luminosity on the Mid-Infrared 
Spectral Energy Distributions of Radio-Quiet Quasars}

\author{S.\ C.~Gallagher,\altaffilmark{1}
G.\ T.~Richards,\altaffilmark{2}
M. Lacy,\altaffilmark{3} 
D.\ C. Hines,\altaffilmark{4}
M. Elitzur,\altaffilmark{5}
\&
L.\ J. Storrie-Lombardi\altaffilmark{3} 
}  

\altaffiltext{1}{Department of Physics \& Astronomy, University of
  California -- Los Angeles, 430 Portola Plaza, Box 951547, Los
  Angeles CA, 90095--1547, USA; {\em sgall@astro.ucla.edu}}
\altaffiltext{2}{Department of Physics \& Astronomy, The Johns
  Hopkins University, 3400 North Charles Street, Baltimore, MD
  21218--2628, USA}
\altaffiltext{3}{\spitzer\ Science Center, Caltech, Mail Code 220--6,
  Pasadena, CA 91125, USA}
\altaffiltext{4}{Space Science Institute, 4750 Walnut Street, Suite
  205, Boulder, CO 80301, USA}
\altaffiltext{5}{Department of Physics \& Astronomy, 600 Rose Street,
University of Kentucky, Lexington, KY 40506--0055, USA}

\begin{abstract}
We present an analysis of the effects of luminosity on the shape of
the mid-infrared spectral energy distributions (SEDs) of 234
radio-quiet quasars originally presented by Richards et al.  In
quasars without evident dust extinction, the spectrally integrated
optical and infrared luminosities are linearly correlated over nearly
three decades in luminosity.  We find a significant ($\gtrsim99.99\%$
confidence) correlation between the 1.8--8.0\micron\ spectral index
and infrared luminosity that indicates an enhancement of the
mid-infrared continuum with increasing luminosity.  Coupled with
strong evidence for spectral curvature in more luminous quasars, we
conclude this trend is likely a manifestation of the `near-infrared
(3--5\micron) bump' noticed in earlier quasar SED surveys. The
strength of this feature is indicative of the contribution of emission
from the hottest ($\gtrsim1000$~K) dust to the mid-infrared spectrum;
higher luminosity quasars tend to show more hot dust emission.
Finally, the comparable distribution of bolometric corrections from
the monochromatic 3\micron\ luminosity as well as its lack of
sensitivity to dust extinction as compared to the standard bolometric
correction from $\nu L_{\rm 5100\AA}$ suggests that the former may be
a more robust indicator of bolometric quasar luminosity.  The close
link between the power in the mid-infrared and optical and the effect
of luminosity on the shape of the mid-infrared continuum indicate that
considering mid-infrared emission independent of the properties of the
quasar itself is inadequate for understanding the parsec-scale quasar
environment.
\end{abstract}

\section{Introduction}
\label{sec:intro}

Empirically, quasars are remarkable for spectral energy distributions
(SEDs) with substantial power spanning many decades of frequency from
the radio through the hard X-ray \citep[e.g.,][]{ElvisEtal1994}.  This SED
encodes (though perhaps not unambiguously) 
the fundamental physical properties of the accreting black
hole system such as the mass, spin, and accretion rate as well as
random effects due to orientation; determining the mapping from the
physical properties of interest to their spectral signatures is the
ultimate goal of quasar SED studies.

The accretion flow near the black hole generates the vast majority of
the continuum emission from the near-infrared through the hard X-rays,
while infrared continuum light longward of $\sim1$\,\micron\ is
generally attributed to a parsec-scale, cold, and dusty region that is
heated by and reprocesses the higher energy, direct continuum and
serves to obscure the inner accretion system and broad-line region in
type~2 (here defined as narrow-line) quasars.  While the community
generically refers to this medium as a `torus', this convenient
picture merely accounts for the empirical evidence for axisymmetric
dust obscuration without providing a coherent physical description of
this structure (see \altcite{KoKa1994} and \altcite{ElSh2006} for
possible dynamical models).  Regardless of the specific model, the
quasar torus might be considered the interface between a quasar and
its host galaxy, a reservoir for gas to fuel the black hole and
potentially a site of star formation.  Understanding the nature of the
torus is a primary aim of infrared quasar studies.

\citet[][hereafter R06]{ric+06} recently presented 24\,\micron\ through
$u$-band photometric data of the largest sample to date of type~1
(broad-line) quasars.  The resulting 259 SEDs, supplemented with
archival VLA, {\em GALEX}, and {\em ROSAT} data as available, span a
bolometric luminosity range of approximately three decades.  Overall, SEDs
of type~1 quasars are remarkably similar out to $z=6$
\citep[e.g.,][]{ohad+06,hines+06,jiang+06}.  However, in a qualitative
comparison of two composite SEDs made from higher and lower luminosity
objects, R06 noted subtle differences indicating that luminosity
affected overall SED shapes, particularly in the mid-infrared
(1.3--8\,\micron) as shown in Figure~\ref{fig:sed_zoom}.
Such differences in the mid-infrared spectra of quasars with higher
mid-infrared fluxes were noted previously in the SEDs of a smaller sample of
35 optically selected quasars by \citet{hatz+05}, and even earlier in
an SED study of 29 Seyfert galaxies \citep{EdMa1986}.

Luminosity is already known to affect some quasar emission properties.
For example, the Baldwin Effect characterizes the decreasing
equivalent widths of ultraviolet emission lines with increasing
ultraviolet continuum luminosity
\citep[e.g.,][]{baldwin,ric+02,shang+03}.  In addition, the maximum
velocity of outflows seen in \CIV\ absorption increases with
ultraviolet luminosity \citep{LaoBra2002}. Across broader spectral
regions, the X-ray-to-optical flux ratio is known to decrease with
increasing ultraviolet luminosity
\citep[e.g.,][]{AvTa1986,Strateva+2005}.  Finally, and perhaps most
relevant for the current study, the ratio of type~1 to type~2 quasars
is found to increase with luminosity, as found in X-ray
\citep[e.g.,][]{ueda+03,akylas+06} and narrow emission-line surveys
\citep{hao+05}. Overall, these empirical relationships indicate that
luminosity affects the quasar system in some fashion.

In this paper, we investigate potential SED differences of type~1
quasars as a function of luminosity, focusing in particular on the
rest-frame mid-infrared and optical/UV emission with the goal of
understanding how the quasar affects its immediate environment.
Throughout we assume $\Lambda$-CDM cosmology with $\Omega_{\rm
M}=0.3$, $\Omega_{\Lambda}=0.7$, and $H_0=70$\kms\,Mpc$^{-1}$
\citep{sperg03,sperg06}.

\section{Analysis and Results}
\label{sec:analysis}

We take as the starting point for our analysis the photometric SED
data from Tables 1 and 2 of R06, focusing on the spectral regions with
complete coverage: the observed-frame optical and mid-infrared.  The
optical coverage is from the Sloan Digital Sky Survey $ugriz$
photometric data \citep{sdss_ref} corrected for Galactic extinction.
The mid-infrared data are from {\em Spitzer} IRAC 3.6, 4.5, 5.8, and
8.0\micron\ (full widths at half maxima [FWHM] of the spectral
responses are 0.7, 1.0, 1.4, and 2.9\micron, respectively) and MIPS
\mbox{24\micron} (FWHM=5.3\micron) imaging.  There has
been no attempt to remove host galaxy light, which is likely to
contribute most significantly to the SED at rest-frame $\sim1$\micron.

R06 included all SDSS quasars with publicly available \spitzer\
coverage as of the Data Release 3 quasar catalog of \citet{dr3_qso}.
The SDSS is quite complete to all but the reddest quasars to $i=19.1$
\citep{ric+06b}; all SDSS quasars were detected by \spitzer.  From the
entire R06 sample of 259 quasars, those that are known to be radio
luminous ($\log(L_{\rm R}) > 32$ erg\persec~Hz$^{-1}$ at rest-frame
20~cm assuming $\alpha_{\rm \nu,R}=-1$) or have no 24\,\micron\
photometry have been eliminated.  Note that the majority of R06
quasars (229/259) have only radio upper limits.  Of these 229, 99 do
not have sensitive enough flux limits to rule out high radio
luminosity.  The trimmed sample comprises \total\ quasars with
$z=0.14$--5.21, $\bar{z}=1.54\pm0.92$, $i=16.8$--20.4,
$\bar{i}=18.9\pm0.7$, $M_{i}=$--22.0 to --28.2, and
$\bar{M_{i}}=-25.6\pm1.6$.

Throughout this section, when we refer to a linear fit, we are using
the ordinary least-squares bisector method to determine the slope and
intercept of a line fitted to (usually logarithmic) data values.
\citet{isobe+90} recommend this method for determining the functional
relationship between ordinate and abscissa values when the errors are
not well defined or insignificant.  Errors quoted for the slopes are
the standard deviations of the best-fitting values.

\subsection{Relative Integrated Infrared and Optical Luminosities}
\label{sec:lopt}

As a first step, we investigated whether the relative power in the
infrared vs. the optical varies with luminosity.  As mid-infrared
emission and optical photons are generated in physically distinct
regions, there is no reason a priori why the relative power should (or
should not) be affected by luminosity.  Figure~\ref{fig:lumplots}a
shows the logarithm of the ratio of infrared and optical luminosities
vs. the logarithm of the optical luminosity; the quantity \lir--\lopt\
may be considered a parameterization of the reprocessed to direct
accretion-powered luminosity.  An alternate depiction of the same
data, \lir\ vs. \lopt\ is presented in Figure~\ref{fig:lumplots}b.  We
use \lir\ and \lopt\ to indicate $\log(L_{\rm opt})$ and $\log(L_{\rm
ir})$, respectively, with units of \lumin; Both values are taken from
Table~1 of R06.  The first, \lir, is the luminosity integrated from
$\log(\nu)=12.48$--14.48 Hz (1--100\micron); \lopt\ is integrated from
$\log(\nu)=14.48$--15.48 Hz (0.1--1\micron).  Power-law interpolation
was used on the monochromatic luminosities calculated from the
tabulated flux densities.

From visual inspection, there is a clear trend with
increasing optical luminosity for quasars to have relatively more
power in the optical, and a linear fit gives a slope of $-0.51\pm0.06$
(1$\sigma$ uncertainty). Non-parametric statistical tests of the
bivariate data measuring Spearman's $\rho$=--0.288 and Kendall's
$\tau$=--0.194 indicate that the probability of obtaining these values
if the parameters are not correlated is $<1\times10^{-5}$; the
statistical values and probabilities are listed in
Table~\ref{tab:stats}.

Given that \lopt\ characterizes direct emission from the accretion
disk, it may be considered in some sense a truer measure of overall
quasar power than the indirect and reprocessed infrared light. A
straightforward interpretation of this significant trend would be that
a smaller fraction of the accretion-powered luminosity is reprocessed
and emitted in the mid-infrared in more luminous quasars.  However,
optical photons are subject to dust reddening and extinction.  The
integrated optical luminosity measured by the observer would be
reduced by extinction while \lir--\lopt\ would increase in the same
sense as the correlation.  Therefore, if dust extinction is prevalent,
it could induce the correlation seen in Figure~\ref{fig:lumplots}a.
As the small grains responsible for moderate optical extinction will
not affect the infrared power (and may not in any case be along the
same line of sight), \lir\ may be a more robust measure of the total
quasar power. Indeed, we perform the same statistical tests on
\lir--\lopt\ vs. \lir\ and find that these two quantities are not
correlated. The scatter about the mean is also roughly the same across
almost three decades in \lir.

\subsection{Comparison of Optically Blue and Red Quasars}
\label{sec:reddening}

The values of \lir--\lopt\ evident in Figure~\ref{fig:lumplots}a span
roughly one decade from --0.5 to 0.5.  As the apparent trend of
\lir--\lopt\ vs. \lopt\ plotted in Figure~\ref{fig:lumplots}a may be
induced by extinction, we investigate if extinction could also
contribute to the width of the spread in \lir--\lopt.  To simply
parameterize the color of the optical continuum, a power-law model
($l_\nu\propto\nu^{\alpha_{\nu{\rm, opt}}}$) was fit to the SDSS
photometric data between rest-frame 1200--5000\,\AA. This region of
the spectrum is typically dominated by the quasar, and by inspection
of the continuum fits compared to the data (2--5 points in this
region); a power-law model is appropriate for this purpose.  For
illustration, three sample SEDs are shown in Figure~\ref{fig:sample}.
From analysis of the SDSS composite quasar spectrum, the host galaxy
contribution reddens the optical continuum slope for
$\lambda>5000$\,\AA\
\citep{VandenBerk2001}.  Wavelengths $<1200$\,\AA\ are to be avoided
as the intervening Ly$\alpha$ forest reduces the flux significantly.

A histogram of \aopt\ values is shown in Figure~\ref{fig:hists}a.  At
the blue (more positive) end, the distribution is roughly gaussian; a
long tail extends to the red (negative) end.  The qualitative shape of
this \aopt\ distribution is consistent with that of \delg, the SDSS
$g-i$ color minus the mode of the $g-i$ color of quasars at the same
redshift \citep{ric+03}.  The very red tail in the \delg\ distribution
is associated with continuum dust reddening, though intrinsically red
quasars may contribute somewhat.  To illustrate the effect of optical
extinction on Figures~\ref{fig:lumplots}a, a reddening vector has been
drawn.  The vector reflects the decrease in \lopt\ of $A_{\rm g}=0.1$
magnitudes of extinction, assuming Small Magellanic Cloud (SMC)
reddening (as appropriate for quasars; e.g., \altcite{hopkins+04}) of
an
\aopt$=-0.49$ optical continuum integrated from 1000--10,000\,\AA\
(following R06).  

We use the median \aopt=$-0.49$ to divide the sample into `optically
red' (\aopt\,$<-0.49$) and `optically blue' (\aopt\,$\ge-0.49$)
groups.  A comparison of the histograms of
\lir--\lopt\ for the groups (Figure~\ref{fig:hists}b) shows an
offset.  A student's t-test of the two populations indicates that the
probability that they have the same mean is $<1\times10^{-6}$.  This
mean reduction in \lir--\lopt\ for the red quasars is consistent with
the expectation from extinction.  In fact, within the optically red
quasars, the subset of clearly dust-reddened (\aopt$<-0.70$) quasars
show the most extreme values of \lir--\lopt.

Therefore, we conclude that optical extinction makes a significant
contribution to the width in the \lir--\lopt\ distribution, and likely
induces the correlation of \lir--\lopt\ vs. \lopt.  In fact, this
correlation disappears when only blue quasars are considered (see
Table~\ref{tab:stats}).  For optically blue quasars, arguably the
least likely to suffer from dust reddening, the data are consistent with a
constant ratio of infrared to optical luminosity over approximately
three decades in luminosity: $L_{\rm ir}\propto L_{\rm opt}$ (see
Figure~\ref{fig:lumplots}b).

\subsection{Dependence of the Mid-Infrared SED on Luminosity}
\label{sec:midir_spec}

A close-up of the composite SEDs from R06 in the spectral region from
1--12\,\micron\ reveals distinctions between the luminous and faint
quasar SEDs (Figure~\ref{fig:sed_zoom}).  When the SEDs are all
normalized to 1.3\,\micron, the luminous SED shows enhanced flux in
the mid-infrared spectral region relative to the mean and faint
composites.  The three SEDs then almost meet again for
$\lambda\gtrsim$10\micron.  Note that these composites were
constructed from broad-band photometry, and therefore do not reflect
interesting mid-infrared features such as strong 10\micron\ silicate
emission recently seen in several quasar spectra
\citep[e.g.,][]{hao+05b,sieb+05}.

To investigate if these differences are evident trends in all of the
data rather than just in the composites, power-law fits have been
performed to the \spitzer\ data in the rest-frame 1.8--8\,\micron\
region.  As available, 15\micron\ {\em ISO} and $JHK$ photometry were
also incorporated.  This spectral region was chosen to maximize the
photometric coverage while reducing to the extent possible
contamination from the host galaxy (which peaks in the near-infrared)
and silicate emission (at $\sim 10$\,\micron).  Even in logarithmic
units, curvature is evident in this spectral region
(Figure~\ref{fig:sed_zoom}), and so the spectral index \air\
($l_\nu\propto\nu^{\alpha_{\nu{\rm, ir}}}$) should only be considered
a simple parameterization of the data.  Sample SEDs and the
corresponding fits are shown in Figure~\ref{fig:sample}.  The value
of \air\ for eight quasars with $z>3.5$ could not be measured because
the MIPS 24\,\micron\ data point was the only one in the bandpass.

A histogram of the distribution of \air\ values is shown in
Figure~\ref{fig:hists}a.  For reference, the mean and standard
deviation of the \air\ values, $-1.20\pm0.34$, are consistent with the
1--10\micron\ spectral indices measured by \citet{haas+03} of
$-1.3\pm0.3$ for 64 Palomar-Green quasars observed with {\em ISO}.

As anticipated from the composite spectral results, the distribution
of \air\ vs. \lir\ shows a clear trend with increasing \lir\ whereby
the more luminous quasars tend to have steeper (more negative)
spectral indices in the mid-infrared; this is evident in
Figure~\ref{fig:air}. This apparent correlation is supported with the
bivariate Spearman and Kendall tests, which give probabilities that
\air\ is not correlated with \lir\ of $6.9\times10^{-5}$ and
$2.7\times10^{-5}$, respectively.  (We use \lir\ instead of \lopt\ to
avoid the problems of underestimating the luminosity due to
extinction.)

Next, we investigate if quasars are also brighter at 1.8--8\,\micron\
relative to the optical continuum region. We choose to measure the
monochromatic luminosities at 3\,\micron\ and 5000\,\AA\ (\lthree\ and
\lfiveoh\ in logarithmic units) to represent the mid-infrared and
optical, respectively. Three \micron\ is near the middle of the
1.8--8.0\,\micron\ bandpass, has good photometric coverage, and should
have minimal contribution from even a star-forming host galaxy;
5000\,\AA\ marks the red end of the quasar-dominated optical/UV
continuum region.  As seen from the results in Table~\ref{tab:stats},
\lthree--\lfiveoh does not change significantly as \lthree\ increases.
If only optically blue quasars are considered (to reduce the potential
effects of extinction), there is still no significant evidence for any
change in \lthree--\lfiveoh\ for increasing \lthree.  The steepening
of the mid-infrared continuum evident from \air\ vs. \lir\ is much
clearer evidence of mid-infrared changes.

In the course of the previous tests, we visually examined each of the
best-fitting optical and mid-infrared power-law fits in comparison
with all of the SDSS and \spitzer\ photometry (cf.,
Figure~{\ref{fig:sample}). Perhaps surprisingly, up to $z\sim1.2$,
extrapolating a simple power-law fit from the IRAC photometry predicts
quite well $L_{\nu}$ in the MIPS 24\micron\ bandpass. From $z=1.2$--2,
however, the extrapolation typically {\em overpredicts} $L_{\nu}$ in
the MIPS 24\micron\ bandpass. (Beyond $z=2$, the MIPS datapoint is
included in the power-law fit.)  This systematic effect is not because
of a contribution of significant 10\micron\ flux from silicate
emission, which would cause the opposite effect.  Instead, the deficit
of 24\micron\ emission is evidence for curvature in the mid-infrared
spectrum.  This is consistent with the luminous and faint composite
SEDs, which when normalized at 1.3\micron\ almost meet again near
10\micron\ (see Fig.~\ref{fig:sed_zoom} and R06).

To characterize spectral curvature, we calculate the monochromatic
8\micron\ luminosity, \Leight, in two ways.  For the first,
\Leightair, the best-fitting mid-infared power-law spectral index and
normalization are used to extrapolate to rest-frame 8\micron.  For the
second, \Leightsed, photometric datapoints within 0.3 dex of
rest-frame 8\micron\ are used to normalize the R06 composite SED to
measure the 8\micron\ luminosity.  The lowest panel in
Figure~\ref{fig:sample} illustrates the discrepancy between
\Leightsed\ and \Leightair\ seen in some objects.  Quasars without at
least three data points in the 1.8--8,0\micron\ regime and photometry
within 0.3 dex of 8\micron\ were eliminated from this analysis to
mitigate redshift effects, the resulting sample has 146 objects.  As
seen in Figure~\ref{fig:leight}, the strong decrease in
\Leightsed/\Leightair\ with increasing \Lir\ is indicative of
significant curvature in the spectrum (see Table~\ref{tab:stats}).

Overall, we find strong evidence for significant differences in the
mid-infrared spectral regime as a function of luminosity.
Specifically, more luminous quasars show steeper (more negative) values of \air.
Extrapolating \air\ to rest-frame 8\micron\ also overpredicts the
8\micron\ luminosity in luminous quasars, thus demonstrating 
spectral curvature.  Both these results are consistent with the
expectation from the luminous composite shown in
Figure~\ref{fig:sed_zoom}.

\subsection{Investigating Potential Systematic Redshift Effects}
\label{sec:z_effects}

Quasar spectral studies have historically been plagued with
difficulties in distinguishing between luminosity and redshift effects
(see \altcite{steffen+06} for a recent example) particularly as
spectral redshifting causes distinct spectral regions to be sampled
in a given observed-frame bandpass.  

In this study, the evident mid-infrared spectral curvature and the gap
in the \spitzer\ photometric coverage between observed-frame 8 and
24\microns\ could induce apparent systematic differences in the
mid-infrared spectra as a result of redshift.  To examine this
potential problem, we focused on a narrow redshift slice in our
sample, and reran the statistical tests of \air\ vs. \lir.  We choose
the redshift range from $z=1.0$--1.5 because there is still a
significant ($\gtrsim1$~dex) infrared luminosity range and at least 3
IRAC bands sample the rest-frame wavelength range of interest.  This
redshift subsample comprises \subsample\ quasars.  Even for this
significantly smaller sample, the correlation of \air\ with \lir\
remains statistically significant, with both Spearman's and Kendall's
tests giving $<1\times10^{-4}$ probabilities of no correlation.
Furthermore, if the effect were entirely induced by bandpass effects,
\air\ would be expected to correlate significantly with redshift,
which is not the case (see Table~\ref{tab:stats}).  More even
photometric sampling is likely needed to better constrain the shape of
the correlation and certainly to investigate spectral curvature in
greater detail.

\subsection{Comparing Optical and Infrared Spectral Indices}

In the standard understanding of the infrared emission in quasars,
cold, dusty material on parsec scales from the supermassive black hole
is illuminated by and reprocesses accretion disk radiation.  The torus
geometry is typically described as some form of flattened disk-like
configuration, encompassing toroids, open cones, and flared disks.  At
mid-infrared (1--10\micron) wavelengths, the continuum spectrum is
dominated by thermal dust emission, and as long as the highest dust
temperature is fixed (by dust sublimation, for example), the SED is
not sensitive to the spectral shape of the illuminating continuum
regardless of the assumed geometry \citep{IvEl1997}.  Only at
near-infrared ($<3$\micron) wavelengths is the illuminating spectral
shape expected to affect the SED as a result of scattering of direct
continuum photons (Nenkova et al., in prep.).  As seen in
Figure~\ref{fig:alphas}, from these data we find no evidence for any
correlation between \aopt\ and \air, confirming theoretical
expectation.  Even when considering just the optically blue quasars to
reduce the potential masking effects of optical extinction, the two
parameters show no relation.  Therefore, there is no evidence that the
mid-infrared spectral changes are induced by the shape of the
illuminating continuum; this is also consistent with the lack of
notable differences between the composite SEDs constructed by R06 from
optically red and optically blue quasars.

\subsection{Evaluating Mid-Infrared and Optical Bolometric Corrections}

Typically, quasars only have photometric data in a few bandpasses
while the quantity of interest (e.g., for estimating black hole masses
and Eddington accretion rates) is the integrated bolometric
luminosity.  Therefore, finding a single point in the SED from which a
robust estimate of the bolometric luminosity can be made would be
useful.  Given the evidence discussed in \S\ref{sec:lopt} for the
potentially significant effects of dust extinction on \lopt, we
investigate whether a mid-infrared point would be more robust.  Using
the 3\micron\ monochromatic luminosity, $\nu L_{\rm 3 \mu m}$, we
calculated both the overall (to $L_{\rm bol}$) bolometric corrections
(BCs) and the infrared (to $L_{\rm ir}$) BCs.  The value of $L_{\rm
bol}$ is integrated from 100\micron\ to 10~keV; where photometry do
not exist to measure this quantity, the composite SED was used to fill
out the wavelength coverage.  These are also compared to the typical,
5100\AA\ BCs as compiled in R06.  Histograms of the three BC
distributions are shown in Figure~\ref{fig:bc}.  The 3\micron\ BC,
8.59$\pm$3.30 is notably quite comparable to the 5100\AA\ BC,
10.47$\pm$4.14, with fractional standard deviations of 38\%\ and 40\%,
respectively.  The extremely tight and symmetric 3\micron\ infrared BC
indicates that $\nu L_{\rm 3 \mu m}$ is quite a good proxy for $L_{\rm
ir}$.

\section{Discussion}
\label{sec:dis}

The distribution of \lir\ vs. \lopt\ as plotted in
Figure~\ref{fig:lumplots}b is notably quite tight.  As expected in a
paradigm where the direct optical/UV continuum emission powers thermal
emission in the infrared, optical and infrared luminosity increase in
tandem.  While the overall sample of 234 quasars gives $L_{\rm
ir}\propto L^{0.94\pm0.02}$, considering only the blue quasars (the
least likely to be dust-extincted) gives a linear relationship.  This
constant ratio of optical to infrared luminosities over $\sim3$
decades in luminosity suggests that the dust mass (estimated from the
infrared luminosity) radiating in the regime probed by \spitzer\ also
increases approximately linearly with optical luminosity (assuming
similar grain properties and distributions). At the same time, the
inner radius of the dust-emitting region increases as $\sqrt{L_{\rm
UV}}$ as the dust sublimation radius moves out.  Thus the inner wall
of the torus at any luminosity is expected to have a similar
temperature.  It is challenging to extrapolate this point to
inferences about the total infared emitting volume, as for optically
thick emission (expected for $\lambda<10$\micron), the observed
luminosity does not depend only on volume, but also on, for example,
dust covering fraction, self-shadowing, and viewing angle.  Therefore,
it is not clear how our results are related to the interpretation that
torus geometry must change with quasar power, as understood from the
increasing type~1/type~2 ratio with luminosity found empirically in
X-ray \citep[][]{ueda+03,akylas+06} and narrow emission-line surveys
\citep{hao+05}.

Indications that quasar luminosity is affecting the torus structure
are the significant spectral changes in the mid-infrared.
Specifically, we seen an enhancement of mid-infrared emission in more
luminous quasars manifested as a steepening of the 1.8--8.0\micron\
continuum.  As expected from the composite SEDs, this enhancement is
better described as a bump, as an extrapolation from \air\
overpredicts the luminosity at $\sim8$\micron\ in luminous quasars.
In current static torus models, the shape of the mid-infrared
continuum shortward of the 10\micron\ silicate feature is sensitive to
numerous physical inputs, including (1) orientation effects
\citep[e.g.,][]{nenkova+02}, (2) the dust grain properties
\citep[e.g., see Fig. 6 of][]{schart+05}, (3) the clumpiness and
number of emitting clouds (\altcite{DuVanB2005}; Nenkova et al., in
prep.), and (4) the opening angle of the torus \citep[e.g., see
Fig. 20 of][]{fritz+06}.  Our focus on type~1 quasars has reduced the
effect of (1) to the extent possible.  Similarly, given that the inner
wall of the torus is set by the graphite sublimation temperature to be
$T\sim1500$~K, a luminosity dependence of dust grain size,
composition, and distribution is not expected, though detailed
infrared spectroscopy is required to examine this issue empirically.
Regarding (3), current clumpy torus models do not generate explicit
predictions for any luminosity dependence of the mid-infrared
continuum.  Therefore, option (4) seems plausible given the other
evidence for a luminosity-dependent covering fraction of the torus.
However, we are not aware of explicit predictions in the literature
for how a changing opening angle would affect the mid-infrared
continuum that are consistent with our results.

Another possible explanation for the enhanced mid-infrared spectra of
luminous quasars is that a hotter dust component may be contributing
more significantly to their SEDs than in lower luminosity objects. A
striking example of this type of component was recently presented by
\citet{RodMaz06} in the near-infrared spectrum of the narrow-line
Seyfert 1 galaxy, Mrk 1239; they interpreted it as thermal emission
from $T\sim1200$~K graphite grains.  This type of spectral feature,
the so-called `near-infrared bump', has been noted for years
\citep[e.g.,][]{HyAl1982}, and its physical origin was interpreted in
the same fashion by \citet{barvainis+87}.  Similar structure is also
evident in the mean radio-quiet composite spectrum of
\nocite{ElvisEtal1994} Elvis et al. (1994; though not, interestingly,
in the radio-loud composite --- see their Fig.~10), while the more
recent combined SEDs of \citet{hatz+05} also show some hint of the
near-infared bump (see their Fig.~9).  In fact, \citet{EdMa1986} (who
modeled the feature as a parabola peaked at 5.2\micron\ in
$\log(f_{\nu})$ vs. $\log(\nu)$ units) found its strength to be
correlated with luminosity in their study of 29 quasar SEDs.  Given
the IRAC wavelength sampling, at higher redshifts, the data are
probing the shorter wavelengths of the mid-infrared bandpass, and the
presence of a `hot bump' peaking between 3--5\micron\ would manifest
itself as a steepening in \air.  The overprediction of \Leight\ from
extrapolations from the mid-infrared power law further supports this
interpretation.  This conclusion is not necessarily inconsistent with
a `receding torus' paradigm whereby more luminous quasars have tori
covering less of the sky \citep{lawrence91} as a flatter dusty
structure would be less likely to block lines of sight to the hottest
dust.

Finally, we also mention that polyaromatic hydrocarbon (PAH) emission
is a possible contributor to the mid-infrared enhancement in more
luminous quasars.  We consider this unlikely however, as the strongest
PAH features (at 6.2\micron\ and 7.7\micron) are not sampled by the
IRAC data beyond $z\sim0.3$ and do not enter the MIPS 24\micron\
bandpass until $z\sim3$.  Other PAH features in the 2--5\micron\ range
are much weaker in star-forming regions and laboratory spectra
\citep{peeters+04}, though they have not been well-studied in luminous
quasar spectra.

The gross similarity of quasar mid-infrared through optical SEDs is
consistent with the finding of \citet{brown+06} that both mid-infrared
and optical type~1 quasar selection give similar space densities and
luminosity functions for $z=1$--5.  Therefore, though optical dust
extinction may induce a false correlation as in
Figure~\ref{fig:lumplots}a, it is unlikely to distort strongly the
study of type~1 quasars.  Similar to our findings, a lack of variation
of the ratio of far-infrared to optical luminosity as a function of
bolometric luminosity was found by \citet{andreani+99} for a sample of
120 quasars (with many far-infrared upper limits).  However, we
caution on the use of far-infrared ($\gtrsim20$\micron) luminosities
to characterize quasars as star formation in the host galaxy can
contribute significantly to or even dominate the power in this regime.
For example, from the 1--100\micron\ spectral modeling of 36 type~1
quasars with luminosities matched to our sample ($L_{\rm quasar} >
3\times10^{44}$\lumin) presented in \citet{fritz+06}, the range of
modeled quasar contributions to the total 5--1000\micron\ luminosities
ranged from 9--100\%\ with a mean of $62\pm26\%$.  The correlation
between the far-infrared luminosities and 7.7\micron\ PAH emission
seen by \citet{schweitz+06} in a sample of quasars and ultra-luminous
infrared galaxies further supports star-formation as a significant
(and variable) source of far-infrared power in luminous active
galaxies.  However, in typical torus+star formation model fits, the
quasar dominates from $\sim2$ to 10\micron\
\citep[e.g.,][]{vanBDu2003}, even considering some contribution from
PAH emission.  As the integrated \Lir\ values used in this work were
derived from IRAC+24\micron\ MIPS data, they may more accurately
represent the quasar power than infrared luminosities estimated from
longer wavelength data.

\section{Summary and Conclusions}

We have analyzed \total\ radio-quiet quasar SEDs from the sample of
R06 with the goal of investigating possible mid-infrared SED changes
as a function of luminosity.  While we find quasar SEDs overall to be
remarkably stable with luminosity in this wavelength regime, there are
some notable differences.  We detail the following results:

\begin{enumerate}

\item{Over three decades in infrared luminosity, the mean ratio and
  scatter of \Lir/\Lopt\ vs. \Lir\ in quasar SEDs is constant.  Some
  part of the spread in the distribution of \Lir/\Lopt\ is incurred
  from optical dust extinction.  Given that dust extinction can
  significantly reduce \Lopt, the mid-infrared luminosity may be a
  more robust indicator of bolometric luminosity.}

\item{The steepening of the 1.8--8.0\,\micron\ spectral index, \air,
  with luminosity coupled with the spectral curvature
  evident from \Leight\ measurements indicate that more luminous
  quasars show a mid-infrared enhancement consistent with a
  3--5\micron\ bump.  This result confirms the qualitative differences
  noted by R06 between luminous and faint composite quasar SEDs. As
  noted in previous studies, this feature can be attributed to an
  increasing contribution from hot ($T>1000$\,K) dust with increasing
  quasar power \citep{barvainis+87}.}

\item{The overall bolometric correction from the monochromatic 3\micron\
  luminosity, 8.59$\pm$3.30, has a comparable dispersion to the
  typical bolometric correction from $\nu L_{\rm 5100\AA}$ without
  suffering from dust extinction.  The infrared bolometric correction
  from $\nu L_{\rm 3\mu m}$ to $L_{\rm ir}$, $3.44\pm1.68$, shows a very tight and
  symmetric distribution, and is therefore a good measure of
  mid-infrared power.}
\end{enumerate}

Though the first two results point strongly to a change in the
geometry of the mid-infrared emitting region with luminosity, we know
of no predictions from current torus models that are consistent with
these findings.  The notable 1.8--8.0\micron\ spectral changes we have found
therefore furnish strong constraints for future theoretical modeling
of mid-infrared torus emission.

Static models for the torus are ultimately unsatisfying; our results
support the picture where quasars are shaping the environment that
emits in the mid-infrared -- the `torus' is not simply a passive
reprocessor of UV/optical emission.  The dynamical wind
paradigm, where magnetohydrodynamic and radiation pressure lift and
accelerate a dusty wind away from the central engine is therefore more
appealing \citep{KoKa1994,EvKoAr2002,ElSh2006}, and radiation pressure naturally
has a luminosity dependence.  Detailed predictions of mid-infrared
spectra within this framework are a necessary next step, and future
studies of 1--10\micron\ infrared spectra of type~1 quasars matched in
redshift with a range of luminosities will illuminate the nature of
the spectral differences we find.  Comparisons with theoretical models of such data will
enable constraints on physical properties of interest such as the
temperature, structure, and radius of the dusty region to aid in
understanding the role of luminosity in shaping the quasar
environment.

\acknowledgements 
We thank Maia Nenkova, Matt Malkan, and Bev Wills for sharing their expertise.
Support for SCG was provided by NASA through the {\em Spitzer}
Fellowship Program, under award 1256317. 
This research has made use of 
the Sloan Digital Sky Survey (http://www.sdss.org).  

{\it Facilities:} {Spitzer ()}, {Sloan ()} 

\clearpage
\begin{deluxetable}{lrcrc}
\tablecolumns{5}
\tablewidth{0pt}
\tablecaption{Results from Non-Parametric Bivariate Statistical Tests
\label{tab:stats}}
\tablehead{
\colhead{Variables} &
\multicolumn{2}{c}{Spearman} &
\multicolumn{2}{c}{Kendall} \\
\colhead{Independent/Dependent (Sample)\tablenotemark{a}} &
\colhead{$\rho$}        &
\colhead{Prob.\tablenotemark{b}}        &
\colhead{$\tau$}        &
\colhead{Prob.\tablenotemark{b}} 
}
\startdata
\lopt/(\lir--\lopt) (234)    &    $-0.288$   & 7.52e--6  & $-0.194$  &   9.66e--6  \\
\lopt/(\lir--\lopt) (blue; 117)   &$-0.077$   & 4.09e--1  & $-0.052$  &   4.08e--1  \\
\lopt/(\lir--\lopt) (red; 117)   &$-0.382$   & 2.18e--5  & $-0.260$  &   3.24e--5  \\

\lir/(\lir--\lopt) (234)  &    $0.039$  &    5.51e--1    & $0.032$ &    4.66e--1     \\
(\lir--\lopt)/\aopt (234)    &    $-0.422$  &  $<1$e--7	&   $-0.292$ &    $<1$e--7 \\
\lir/\air\         (226)   &    $-0.262$   &  6.89e--5  & $-0.188$  &   2.67e--5 \\
\lir/\air\ ($z=1.0$--1.5; 55) & $-0.508$   &  7.48e--5  & $-0.375$   &  5.36e--5 \\
$z$/\air\  (226)           &    $-0.058$   &  3.88e--1  & $-0.061$  &   1.74e--1 \\
\aopt/\air\ (226)    &            0.020    &  7.64e--1  & 0.013     &   7.67e--1 \\
\aopt/\air\ (blue; 113) & 0.126  &  1.84e--1  & 0.088    &    1.68e--1 \\ 
\lthree/(\lthree--\lfiveoh) (234)  &$0.074$  &  2.62e--1  &$0.047$  &    2.85e--1 \\
\lthree/(\lthree--\lfiveoh) (blue; 117)  &$0.235$  &  1.07e--2  &$0.159$  &  1.11e--2 \\
\lir/(\leightsed--\leightair) (146)    & $-0.337$ & 3.17e--05  & $-0.224$  & 5.90e--05 \\
\enddata
\tablenotetext{a}{The sample labels and numbers of data points are
given in parentheses.  `Blue' and `red' refer to \aopt\,$\ge-0.49$ and
\aopt\,$<-0.49$, respectively.}
\tablenotetext{b}{The two-sided probability that the given variables are not
correlated.}
\end{deluxetable}
\clearpage
\begin{figure*}
\plottwo{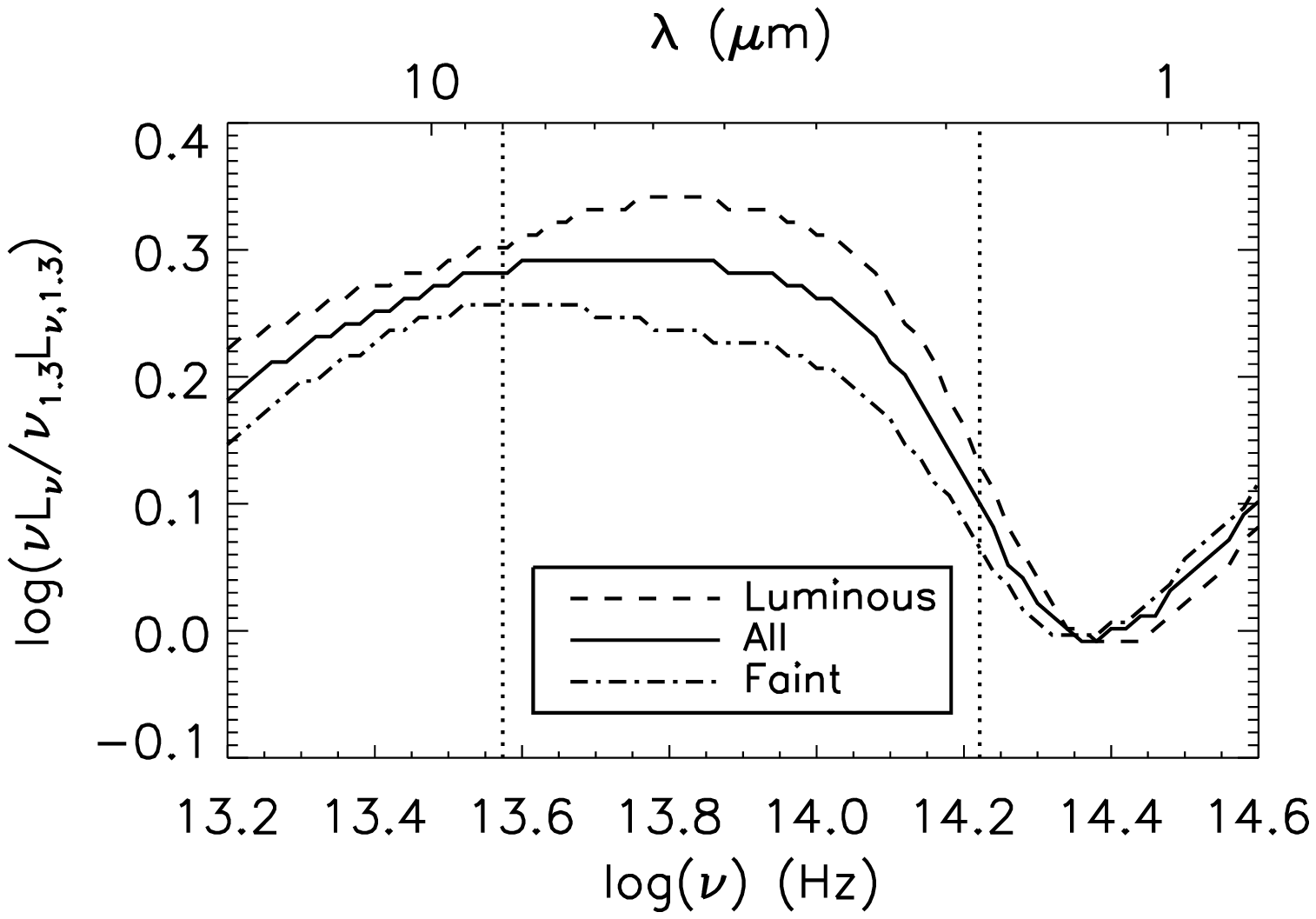}{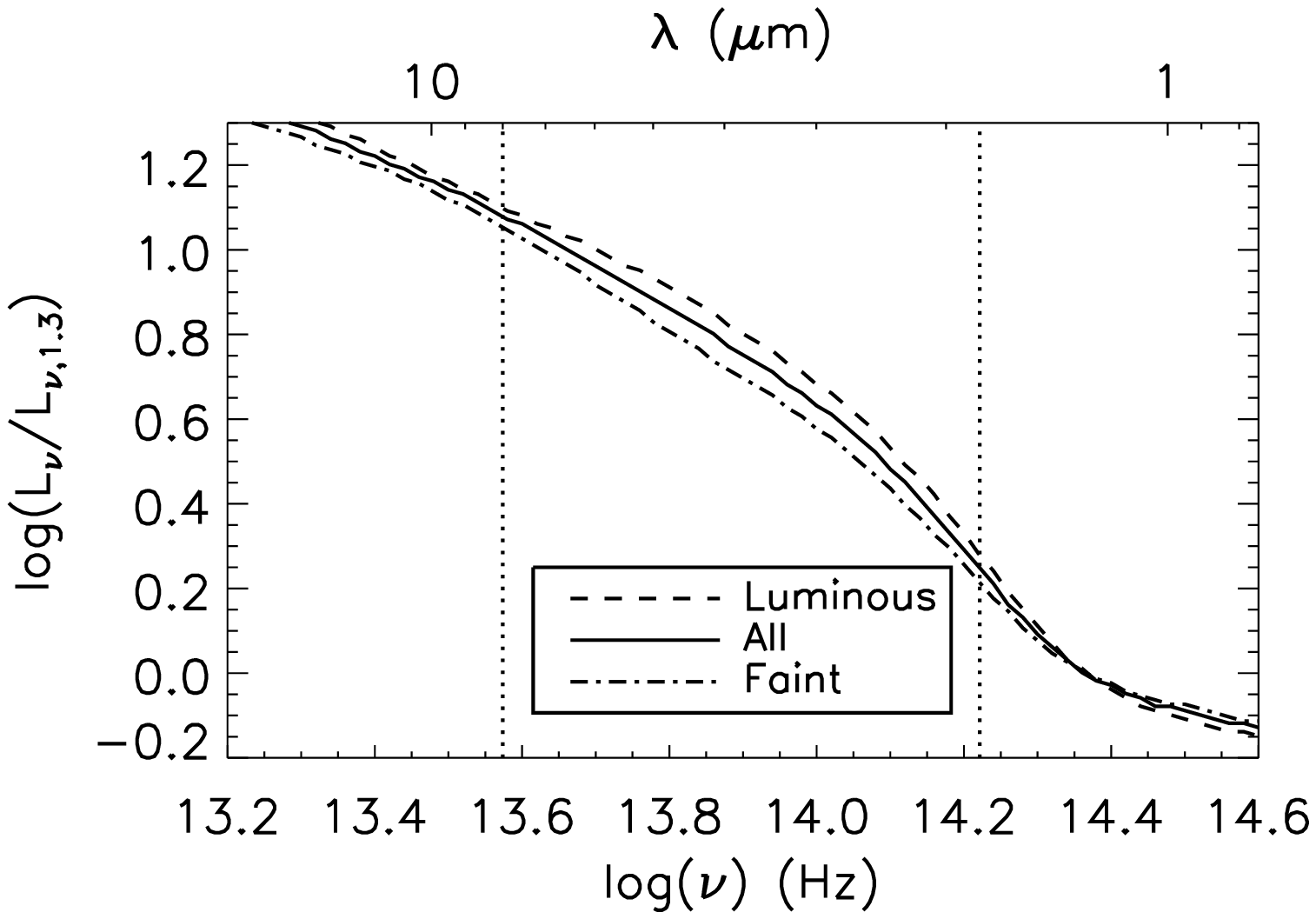}
\caption{ Plots of three SED composites from R06 made from the entire
sample (solid curve), the luminous subset ($\log(L_{\rm opt})>46.02$
\lumin; dashed curve), and the faint subset ($\log(L_{\rm opt})<46.02$
\lumin; dot-dashed curve) in units of {\bf (left)} $\nu L_\nu$, to
emphasize the difference in spectral shapes, and {\bf (right)}
$L_\nu$, to show the spectral region as we model it.  The composite
SEDs have been normalized to match at 1.3\micron.  The SED resolution
is a convolution of the IRAC (FWHM=0.7, 1.0, 1.4, and 2.9\micron\ at
3.6, 4.5, 5.8, and 8.0\micron, respectively) and 24\micron\ MIPS
(FWHM=5.3\micron) filter functions with the redshift distribution.
Statistical subtraction of the host galaxy light is included; for
detailed information on the composite construction see R06. For
reference, 8 and 1.8\micron\ are indicated with vertical dotted lines.
\label{fig:sed_zoom}
}
\end{figure*}
\begin{figure*}
\plottwo{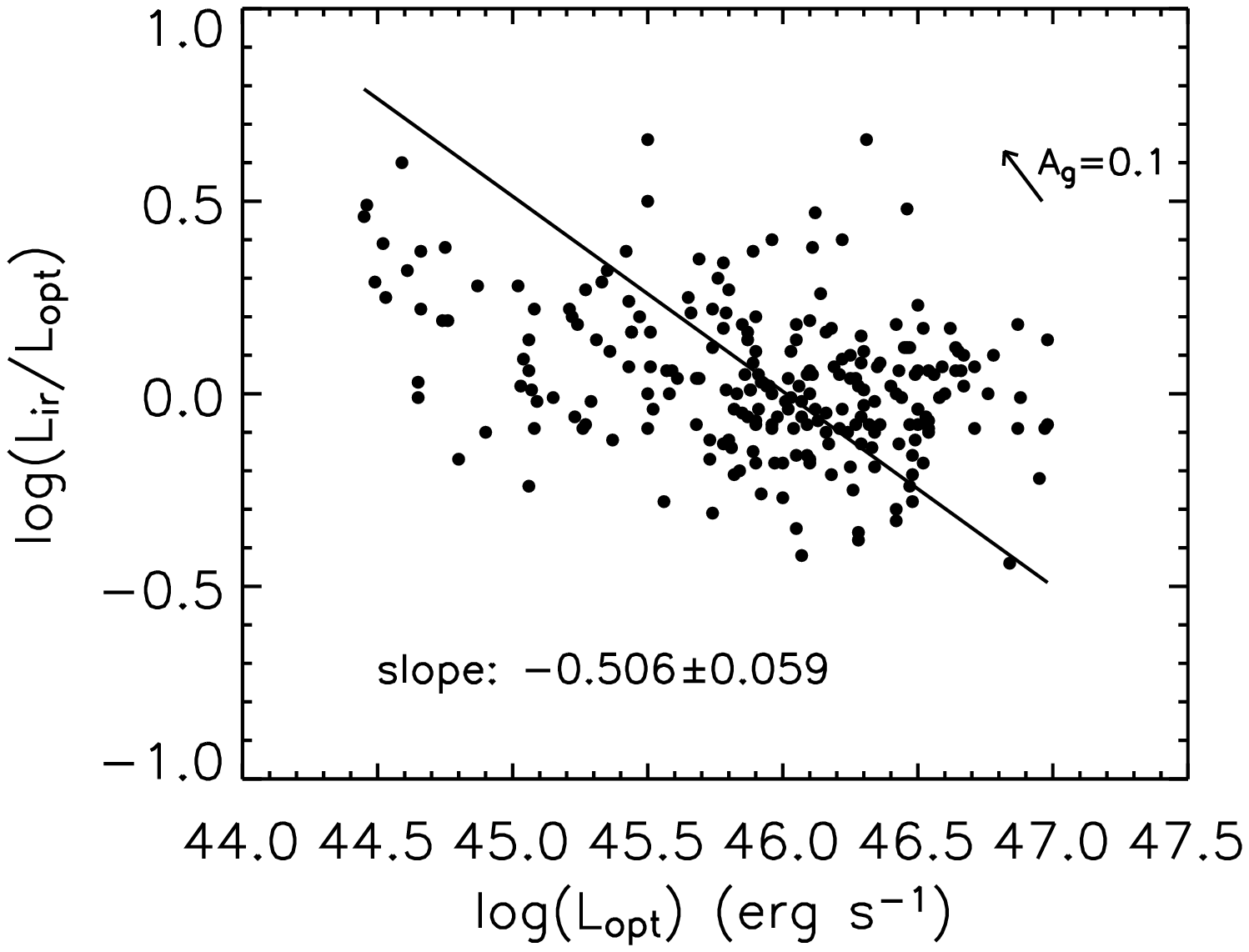}{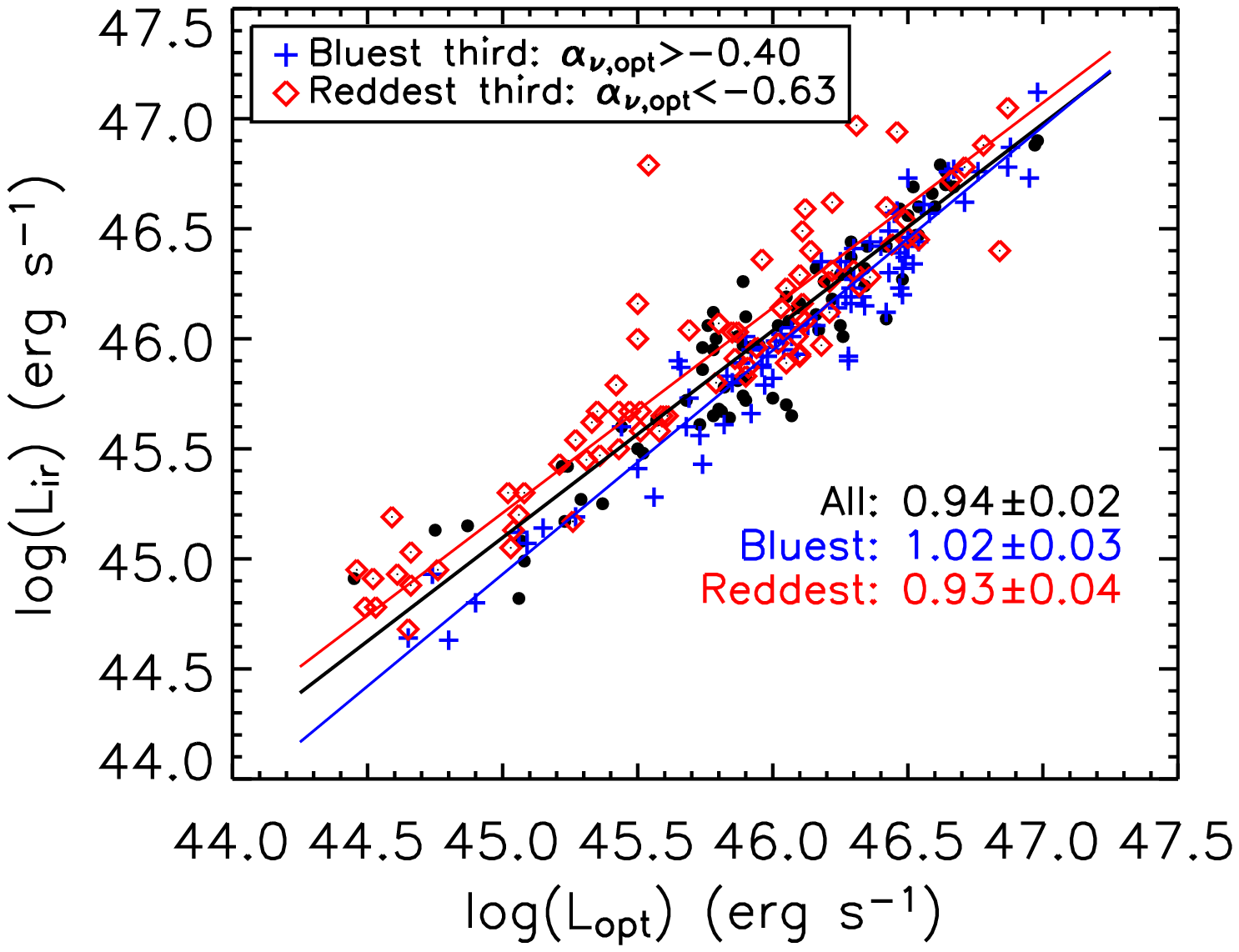}
\caption{{\bf (Left)} Plot of log(\Lir/\Lopt) vs. log(\Lopt) for the
\total\ SED quasars.  The solid line represents a linear fit to the
plotted data; the corresponding slope and standard deviation are given
in the figure.  The labelled vector indicates the magnitude and
direction of the effect of 0.1 magnitudes of $g$-band extinction for
SMC reddening of a power-law continuum with the median
\aopt$=-0.49$. {\bf (Right)} Plot of log(\Lir) vs. log(\Lopt) for the
same sample.  The optically bluest and reddest thirds are indicated
with blue crosses and red diamonds, respectively.  Linear fits to all
the data as well as the bluest and reddest thirds are drawn in solid
lines; the best-fitting slopes are marked on the figure.  While the
lefthand panel indicates a significant correlation between the ratio
of optical to infrared luminosity and the optical luminosity, an
examination of the optically blue data imply that the data are
consistent with $L_{\rm ir}\propto L_{\rm opt}$ if extinction is taken
into account.
\label{fig:lumplots}
}
\end{figure*}
\begin{figure*}
\plotone{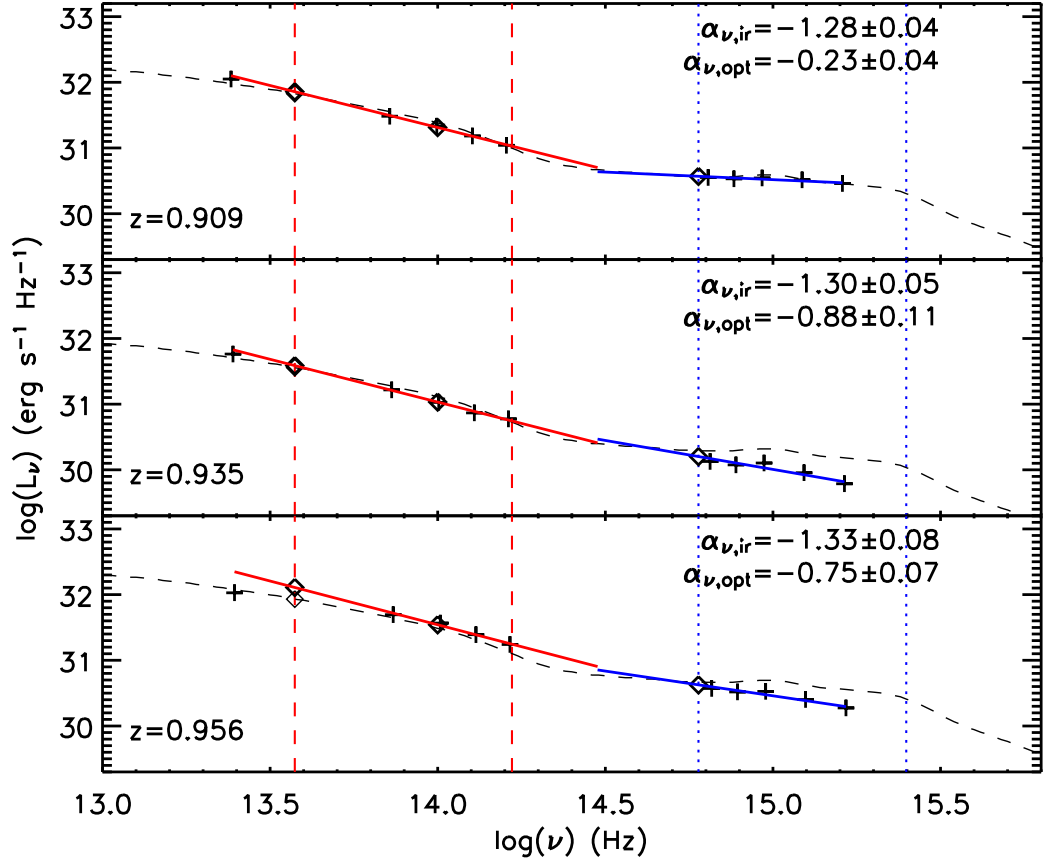}
\caption{Sample SEDs illustrating the data analysis process; the
  abscissa is rest-frame frequency.  Data are labeled with crosses,
  open thick diamonds indicate the computed quantities (from left to
  right) of \Leightair, \Lthree, and \Lfiveoh\ (see
  \S\ref{sec:midir_spec} for definitions).  The thin dashed curve
  shows the mean composite SED from R06 normalized to the photometric
  data within 0.3 dex of rest-frame 8\micron; the thin open diamond
  represents \Leightsed\ (overlapping with \Leightair\ in the top two
  panels).  Red dashed and blue dotted vertical lines mark the
  boundaries of the mid-infrared and UV/optical fitting regimes,
  respectively. Solid lines are the best-fitting power-law models to
  the mid-infrared and optical data; they have been extrapolated to
  cover all of the data.  The top panel is from a quasar with data
  closely matching the composite SED.  The middle example shows a red
  (likely from dust reddening -- note the decreased \Lopt) optical
  continuum, while the bottom panel indicates a measurable discrepancy
  between \Leightsed\ and \Leightair\ indicating mid-infrared spectral
  curvature.
\label{fig:sample}
}
\end{figure*}
\begin{figure*}
\plottwo{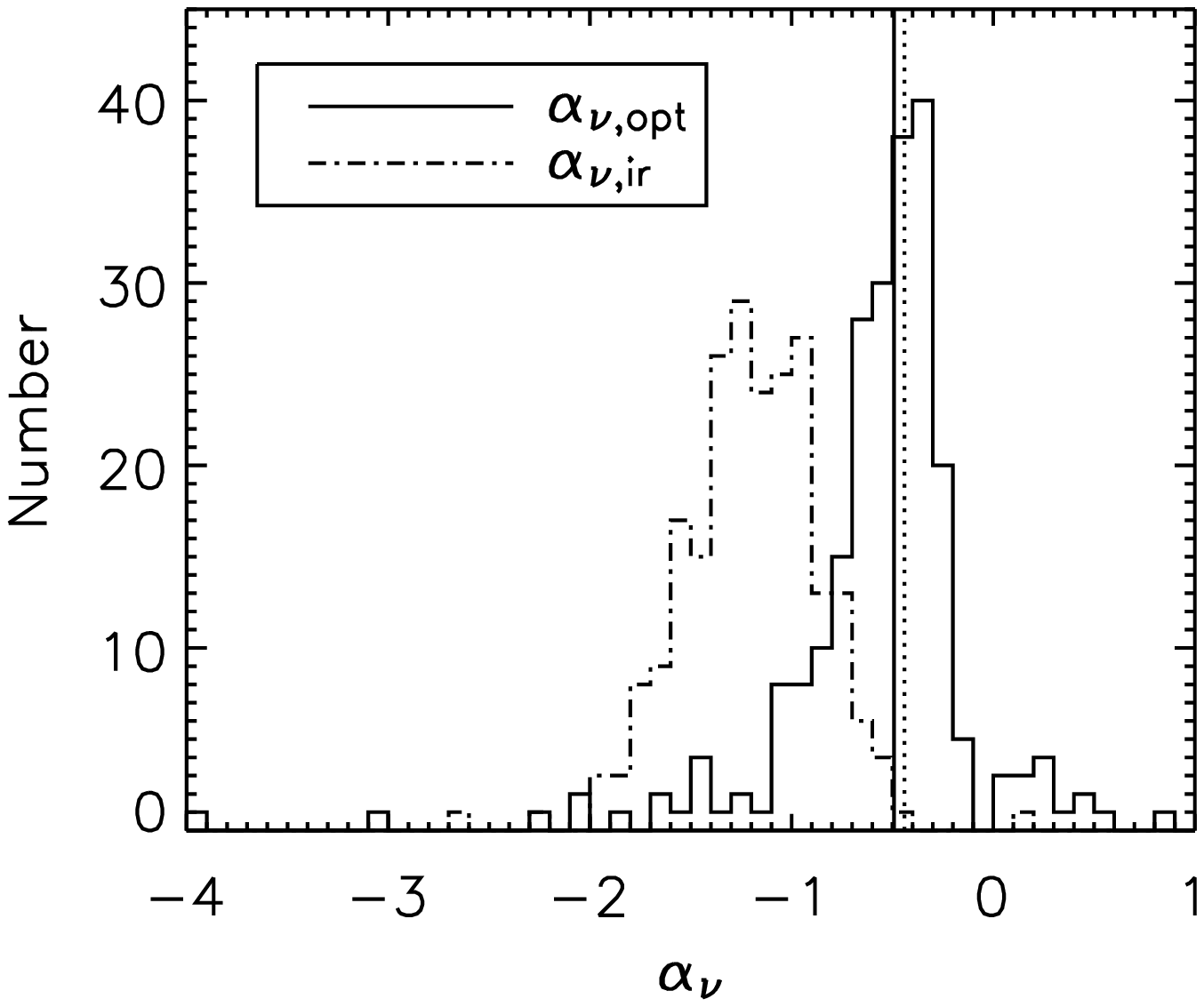}{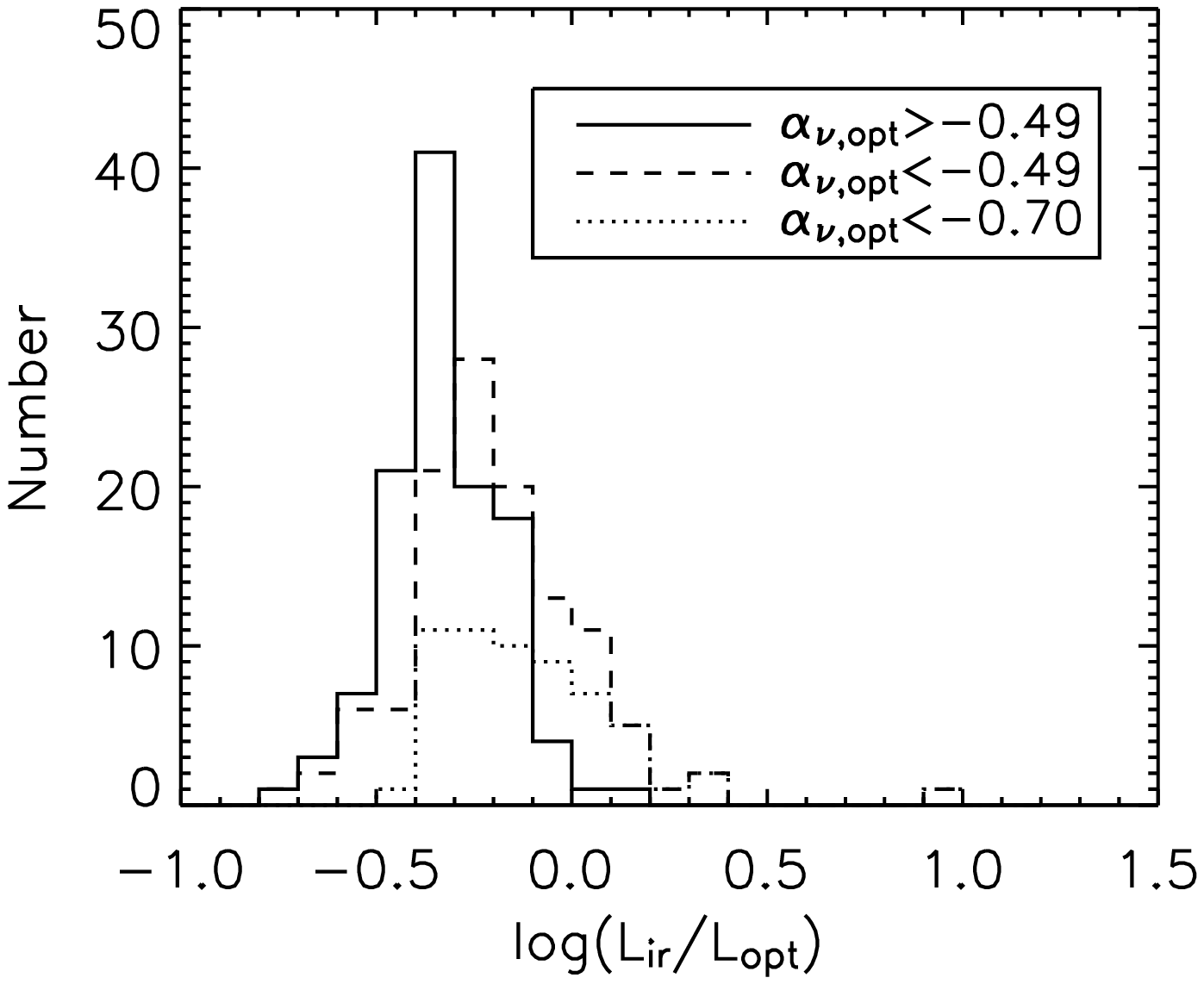}
\caption{ {\bf (Left:)} Histograms of \aopt\ (solid) and \air\
(dot-dashed) for the SED sample. The $\alpha_{\nu}$ values are
determined from power-law fits ($l_{\nu}\propto\nu^{\alpha_{\nu}}$) to
the photometric data; bluer continua are to the right. The median
value for \aopt\ of --0.49 is indicated with a solid vertical line.
For reference, \aopt=--0.44, the best-fit value to the continuum of
the composite SDSS quasar spectrum is also plotted (dotted vertical
line; \citealt{VandenBerk2001}).  {\bf (Right:)} Histograms of the
distribution of log(\Lir/\lopt) for the optically blue (\aopt$\ge-0.49$;
solid), red (\aopt$<-0.49$; dashed) quasar populations.  Quasars that
are very likely to be dust-reddened (\aopt$<-0.70$; dotted) are also
shown. The difference between these histograms indicates that at least
some of the spread in the overall log(\Lir/\Lopt) distribution is an
artifact of optical extinction that reduces \Lopt.
\label{fig:hists}
}
\end{figure*}
\begin{figure*}
\plotone{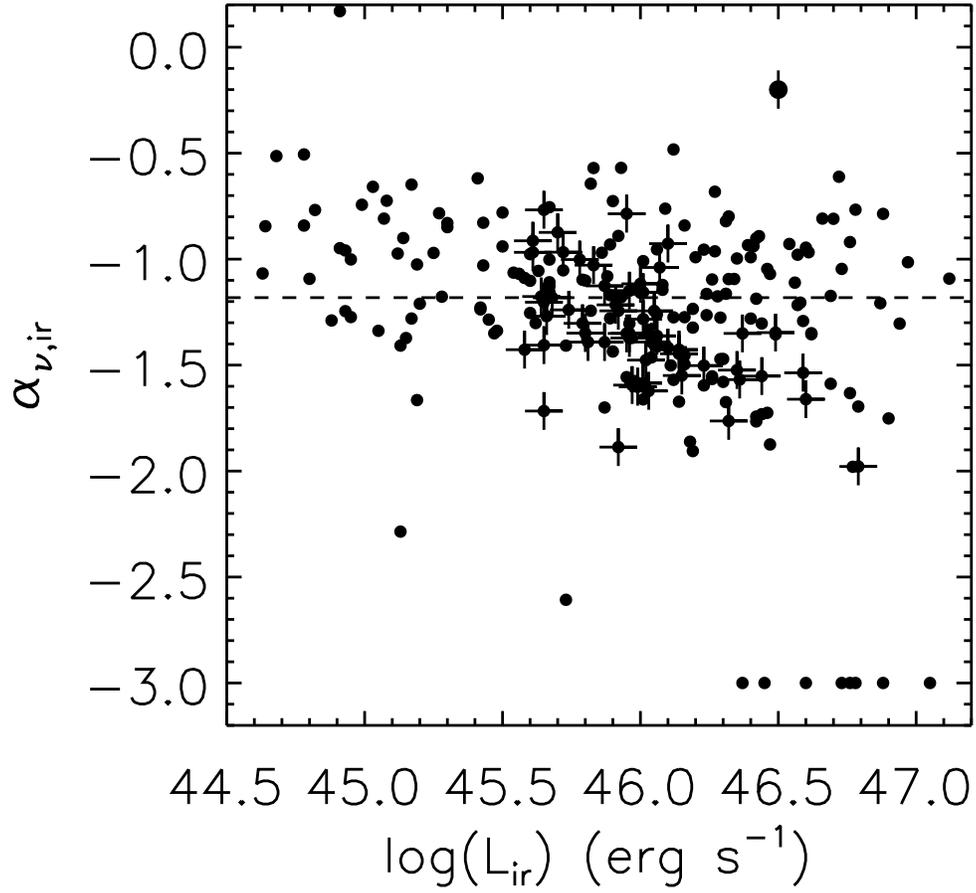}
\caption{ The infrared spectral index, \air, vs. log(\Lir) for the SED
  sample (solid dots); the subset of the SED sample with $z=1.0$--1.5
  is also marked with crosses. The parameter \air\ is determined from
  power-law fits ($l_{\nu}\propto\nu^{\alpha_{\nu{\rm ,ir}}}$) to the
  \spitzer\ photometry between rest-frame 1.8--8.0\,\micron; the eight
  quasars at $z>3.5$ (plotted at \air=--3.0) only had one photometric
  data point in the bandpass; they are not included in the analysis.
  For reference, a horizontal dashed line marks the median \air\
  value; for higher values of \Lir\ a larger fraction of quasars lie
  below the median.  The median uncertainty in \air\ from fitting a
  power-law model to the photometry is indicated as a vertical
  errorbar on a large, filled circle in the upper right corner of the
  plot.
\label{fig:air}
}
\end{figure*}
\begin{figure*}
\plotone{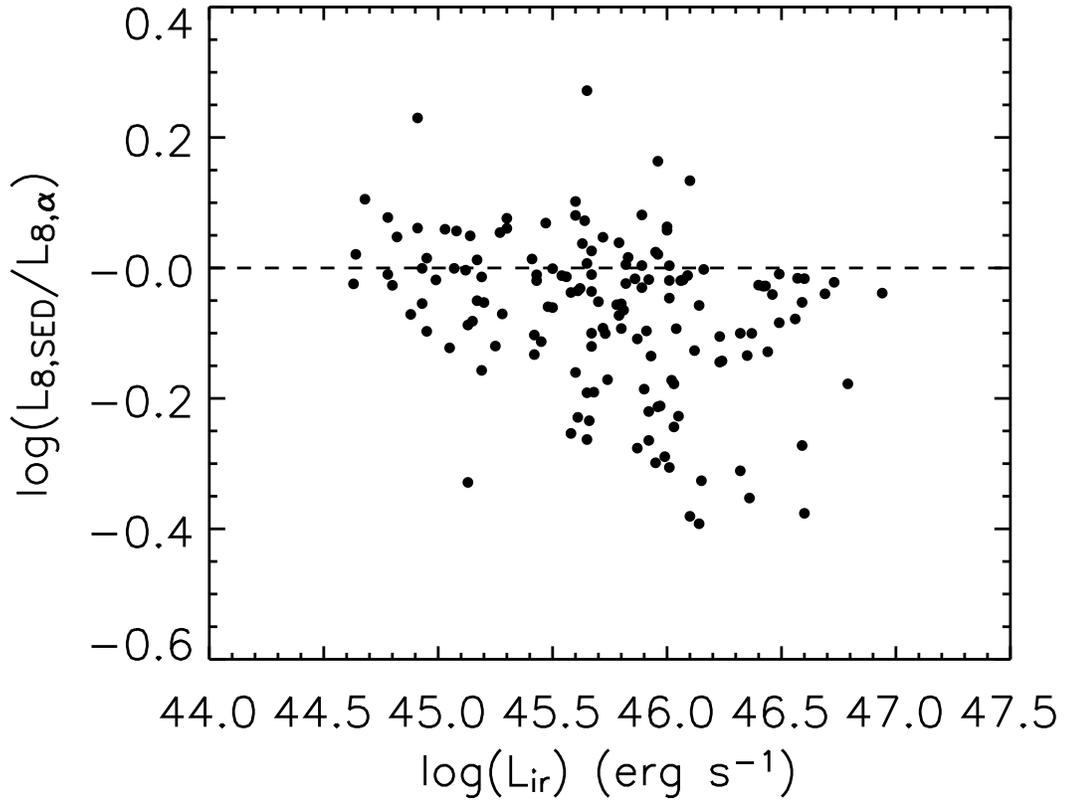}
\caption{The logarithmic ratio of \Leightsed\ (\Leight\ determined
from the R06 mean composite SED normalized to photometry within 0.3
dex of rest-frame 8\micron) to \Leightair\ (\Leight\ calculated by
extrapolating to 8\micron\ from the mid-infrared power-law model)
versus log(\Lir).  For reference, a horizontal dashed lines marks a
ratio of unity.  The strong decrease in \Leightsed/\Leightair\ with
increasing \Lir\ indicates increasing spectral curvature at higher
luminosities.  
\label{fig:leight}
}
\end{figure*}
\begin{figure*}
\plotone{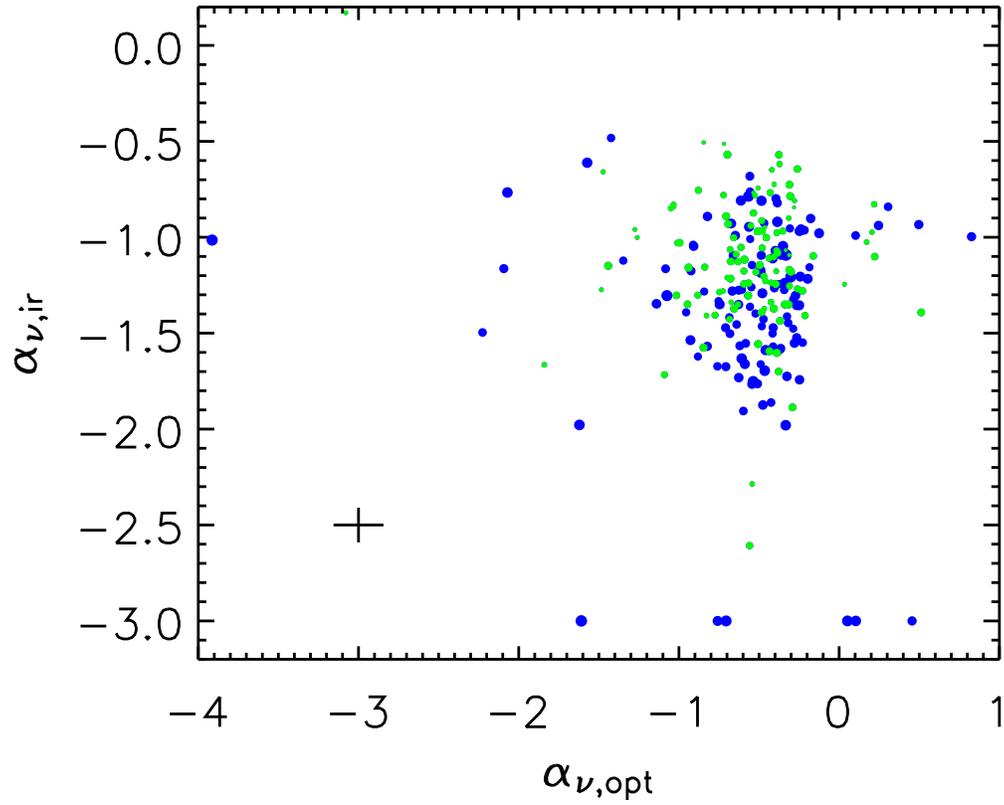}
\caption{ The infrared spectral index, \air, versus the optical
spectral index, \aopt, for the entire SED sample.  Quasars with only
one photometric data point in the 1.8--8.0\micron\ bandpass are
plotted at \air\,$=-3.0$. The symbol sizes have been scaled to
indicate the relative \Lir\ of each quasar; the larger symbols
indicate more luminous quasars.  For clarity, quasars with
log(\Lir)$<46.0$\lumin\ (the median of the \Lir\ distribution) have
been colored green; more luminous quasars are blue.  The cross in the
lower left corner of the plot indicates the size of the median errors
for the fitted values of \air\ and \aopt.
\label{fig:alphas}
}
\end{figure*}
\begin{figure*}
\plotone{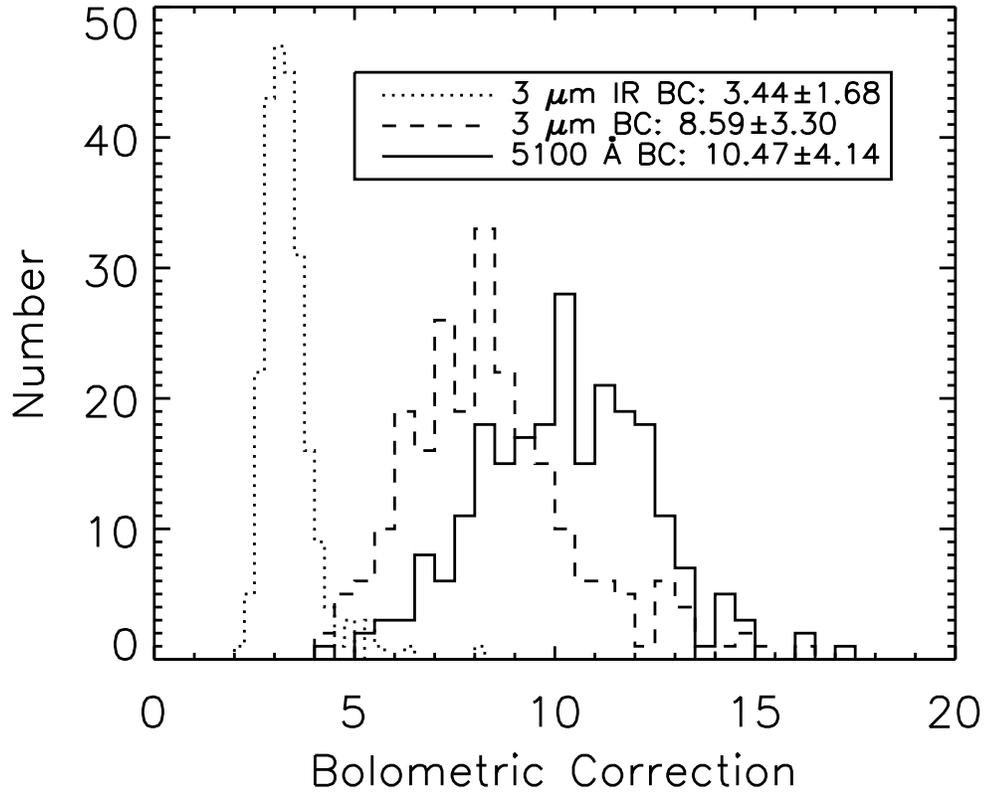}
\caption{ The distributions of bolometric corrections.  The dotted
histogram shows the corrections from $\nu L_{\rm 3\mu m}$ to $L_{\rm
ir}$; the dashed histogram is the corrections from $\nu L_{\rm 3\mu
m}$ to $L_{\rm bol}$, and the solid histogram represents the
bolometric corrections from $\nu L_{\rm 5100\AA}$ to $L_{\rm bol}$ as
tabulated by R06. The legend lists the mean and standard deviation of
the mean for each distribution.  The overall bolometric corrections
from $\nu L_{\rm 3 \mu m}$ show a comparable fractional dispersion to
the standard (5100\AA) bolometric corrections without being subject to
dust extinction effects.
\label{fig:bc}
}
\end{figure*}

\end{document}